\documentclass[prd,preprintnumbers,nofootinbib]{revtex4}
\usepackage{amsfonts}
\usepackage{amssymb}
\usepackage{amsmath}
\usepackage[dvipdfmx]{graphicx}

\begin{document}
\preprint{YITP-16-16}
\preprint{IPMU16-0018}
\title{Ghost inflation and de Sitter entropy}
\author{Sadra Jazayeri\textsuperscript{1}}
\author{Shinji Mukohyama\textsuperscript{2,3}}
\author{Rio Saitou\textsuperscript{4}}
\author{Yota Watanabe\textsuperscript{3,2}}
\affiliation{\textsuperscript{1}School of Astronomy, Institute for Research in Fundamental Sciences (IPM), P.~O.~Box 19395-5531, Tehran, Iran 
}
\affiliation{\textsuperscript{2}Center for Gravitational Physics, Yukawa Institute for Theoretical Physics, Kyoto University, 606-8502,
Kyoto, Japan}
\affiliation{\textsuperscript{3}Kavli Institute for the Physics and Mathematics of the Universe (WPI),
The University of Tokyo Institutes for Advanced Study, The University of Tokyo, Kashiwa, Chiba 277-8583, Japan}
\affiliation{\textsuperscript{4}School of Physics, Huazhong University of Science and Technology, Wuhan 430074, China}

\affiliation{}
\date{\today}

\begin{abstract}
In the setup of ghost condensation model the generalized second law of black hole thermodynamics can be respected under a radiatively stable assumption that couplings between the field responsible for ghost condensate and matter fields such as those in the Standard Model are suppressed by the Planck scale. Since not only black holes but also cosmology are expected to play important roles towards our better understanding of gravity, we consider a cosmological setup to test the theory of ghost condensation. In particular we shall show that the de Sitter entropy bound proposed by Arkani-Hamed, et.al. is satisfied if ghost inflation happened in the early epoch of our universe and if there remains a tiny positive cosmological constant in the future infinity. We then propose a notion of cosmological Page time after inflation. 
\end{abstract}

\maketitle

\section{Introduction}

In the history of quantum mechanics, researches on blackbody radiation and hydrogen atoms played important roles. It is probably cosmology and black holes that are expected to play similar roles towards our better understanding of gravity.

Black holes have properties similar to thermodynamics, known as black hole thermodynamics. Because of the no-hair theorems, stationary black holes have only a small number of parameters while a non-gravitating system in thermodynamic equilibrium are also parametrized by a small number of macroscopic parameters. Similarly to ordinary thermodynamics, the black hole thermodynamics has four laws~\cite{Bardeen:1973gs}: the zeroth law states that the surface gravity $\kappa$ of a stationary black hole is constant over the horizon; the first law relates variation of the mass $M_{\rm bh}$ in the space of stationary black holes to that of the horizon area $A$ as 
\begin{equation}
 dM_{\rm bh} = \frac{\kappa dA}{8\pi G_{\rm N}} + (\mbox{rotation } \& \mbox{ charge terms})\,,
\end{equation}
where $G_{\rm N}$ is the Newton's constant; the second law states that the $A$ does not decrease; and the third law states that $\kappa$ cannot become zero within finite affine time.

From the four laws of black hole thermodynamics it is expected that the horizon area $A$ or its increasing function plays the role of entropy in ordinary thermodynamics~\cite{Bekenstein:1973ur}. It is the Hawking's discovery~\cite{Hawking:1974sw} of thermal radiation from black holes that determined the precise form of the black hole entropy: by identifying the temperature of a black hole to be the temperature of Hawking radiation, 
\begin{equation}
 T = \frac{\hbar\kappa}{2\pi ck_{\rm B}}\,, \label{eqn:T}
\end{equation}
and the energy of the black hole to be $M_{\rm bh}c^2$, where $c$ is the speed of light, one concludes from the first law that the entropy of the black hole should be given by 
\begin{equation}
 S = \frac{k_{\rm B}c^3A}{4\hbar G_{\rm N}}\,, \label{eqn:S}
\end{equation}
up to addition of an arbitrary constant, which can be set to zero by demanding that $S$ vanish for $A=0$. It is also the Hawking radiation that invalidates the second law of black hole thermodynamics: a black hole may lose its mass due to the Hawking radiation and thus the horizon area may decrease. Taking into account the Hawking radiation, the generalized second law then states that the sum of the black hole entropy and the entropy of matter outside the black hole should not decrease. Intriguingly, the formula (\ref{eqn:S}) for the black hole entropy, known as Bekenstein-Hawking formula, involves gravity ($G_{\rm N}$), quantum mechanics ($\hbar$) and statistical mechanics ($k_{\rm B}$). For this reason, many people believe that the black hole entropy should be a key towards our better understanding of gravity. In particular, it is expected that the entropy of a black hole should reflect the number of microscopic states associated with the black hole.

In theories and states that allow for holographic dual descriptions, the generalized second law is expected to be dual to the ordinary thermodynamic second law since a black hole is probably dual to a thermal state. Therefore, if the generalized second law is violated in a certain setup then it probably implies that there is no corresponding holographic dual description in the conventional sense. In the context of ghost condensation, which is the simplest realization of the Higgs mechanism for gravity~\cite{ArkaniHamed:2003uy}, there are some proposals to violate the generalized second law in the literature~\cite{Dubovsky:2006vk,Eling:2007qd,Jacobson:2008yc}. However, it was later shown that if we propose weak enough direct couplings between the field responsible for ghost condensate and the Standard Model (SM) fields or if they have only gravitational interaction through the minimal coupling, then it is possible to prevent the model from violating the generalized second law in a natural effective field theory (EFT) setup~\cite{Mukohyama:2009rk,Mukohyama:2009um}. The point is that under the natural assumption that the field causing ghost condensate is only responsible for the gravity sector, all the interactions between the field for ghost condensate and the SM fields should be suppressed by the Planck scale, $M_{\rm Pl}=1/(8\pi G_{\rm N})^{1/2}$ since such interactions should disappear in the decoupling limit, $M_{\rm Pl}^2\rightarrow \infty$. Once we admit this at the classical level, the natural assumption of the Planckian suppression is not spoiled by radiative corrections since contributions from graviton loops are suppressed by $1/M_{\rm Pl}^2$ at least. A specific interaction that was assumed to exist in~\cite{Mukohyama:2009rk,Mukohyama:2009um} is suppressed by $M^2/M_{\rm Pl}^2$ in the presence of the ghost condensate, where $M$ ($\lesssim 100\,{\rm GeV}$) is the scale of spontaneous Lorentz violation. Because of the strong suppression, the physical processes in those proposals for violation of the generalized second law are too slow to decrease the entropy before the accretion of the ghost condensate to the black hole increases it. We want to emphasize again that it is possible to violate the generalized second law by directly coupling the SM fields to the field causing ghost condensate with strength stronger than gravitational, but as far as the natrual setup is concerned we steer away from such couplings. As already mentioned at the beginning of the present paper, not only black holes but also cosmology are expected to play important roles towards our deep understanding of gravity. In the present paper we thus consider a cosmological setup to test the theory of ghost condensation. In particular, our main focus in the present paper is de Sitter entropy in inflationary universe.

The rest of the paper is organized as follows. In Sec.~\ref{sec:dSentropy} we briefly review the de Sitter entropy bound conjecture proposed by Arkani-Hamed, et.al.~~\cite{ArkaniHamed:2007ky}. We then seek implications of the de Sitter entropy bound on ghost inflation in Sec.~\ref{sec:ghostinflation}. In particular, we show that the de Sitter entropy bound is satisfied in our universe. In Sec.~\ref{sec:pagetime} we propose a notion of cosmological Page time. Sec.~\ref{sec:summary} is devoted to a summary of the paper and some discussions.

\section{de Sitter entropy bound conjecture}
\label{sec:dSentropy}

It has been known that not only black hole horizons but also de Sitter horizons have properties similar to thermodynamics. Indeed, any bifurcating Killing horizon has temperature given by the same formula (\ref{eqn:T}) \cite{Israel:1976ur}. By applying this to the de Sitter case and adopting the natural unit ($c=\hbar=k_{\rm B}=1$), we obtain $T=H/(2\pi)$, where $H$ is the Hubble expansion rate. Also, the Friedmann equation in a slow-roll inflationary epoch is written in the form of the first law with the de Sitter entropy given by the formula (\ref{eqn:S}), where $A$ is understood as the area of the cosmological horizon~\cite{Frolov:2002va}. Again by adopting the natural unit, we obtain $S=\pi/(G_{\rm N}H^2)$.

In quantum gravity, however, it is believed that a de Sitter space is unstable on the Poincar\'{e} recurrence time scale $H^{-1}e^S$~\cite{Goheer:2002vf,Kachru:2003aw}. In the following we thus consider a de Sitter space as a part of slow-roll inflation. For slow-roll inflation driven by a canonical scalar field, we have
\begin{equation}
 \dot{H} = -4\pi G_{\rm N}\dot{\phi}^2\,.
\end{equation}
This is rewritten as
\begin{equation}
 \frac{dS}{dN} = \frac{8\pi^2\dot{\phi}^2}{H^4} \sim \left(\frac{\delta\rho}{\rho}\right)^{-2}\,,
\end{equation}
where $N$ is the number of e-foldings defined by $dN=Hdt$, $\rho$ is the energy density and $\delta\rho$ is perturbation of the energy density estimated as
\begin{equation}
 \frac{\delta\rho}{\rho} \sim \frac{\delta a}{a} \sim H\delta t \sim H\frac{\delta\phi}{\dot{\phi}} \sim \frac{H^2}{\dot{\phi}}\,.
\end{equation}
Excluding eternal inflation, we have $|\delta\rho/\rho|\lesssim 1$ and thus 
\begin{equation}
 0 < \frac{dN}{dS} \lesssim 1\,,
\end{equation}
meaning that the total number of e-folding $N_{\rm tot}$ is bounded from above as 
\begin{equation}
 N_{\rm tot} \lesssim S_{\rm end} - S_{\rm beginning} < S_{\rm end}\,.
  \label{eqn:NtotSend}
\end{equation}
Here, $S_{\rm end}$ and $S_{\rm beginning}$ are the de Sitter entropy at the end and the beginning of inflation, respectively.

The inequality (\ref{eqn:NtotSend}) can be violated if we include the regime of eternal inflation, where $\delta\rho/\rho$ becomes of order unity or larger. However, perturbation generated during the period of eternal inflation is highly nonlinear and would gravitationally collapse to a black hole later at the time of the horizon reentry. Hence there would be no post-inflationary observer who can observe the modes generated during the epoch of eternal inflation. For this reason, while the bound (\ref{eqn:NtotSend}) was derived for non-eternal slow-roll inflation, it can be extended to general slow-roll inflation with or without the eternal epoch by simply replacing $N_{\rm tot}$ with the observable number of e-foldings $N_{\rm obs}$ as
\begin{equation}
 N_{\rm obs} \lesssim S_{\rm end}\,. \label{eqn:NobsSend}
\end{equation}
While the bound (\ref{eqn:NobsSend}) was derived for slow-roll inflation in the above, it was conjectured that the same bound should hold in other models of inflation as well~\cite{ArkaniHamed:2007ky}. This is the de Sitter entropy bound conjecture.

The bound (\ref{eqn:NobsSend}) states that the observable number of different Hubble volumes from a single Hubble patch, $e^{3N_{\rm obs}}$, be less than $e^{3S_{\rm end}}$. If we can trust the effective field theory (EFT) based on the semi-classical approximation, $e^{3N_{\rm obs}}$ is roughly understood as the observable number of independent modes from inflation. Therefore, neglecting $O(1)$ numerical factors, an intuitive interpretation of the de Sitter entropy bound (\ref{eqn:NobsSend}) is that the logarithm of the observable number of independent modes from inflation be bounded from above by $S_{\rm end}$, as far as the EFT is a valid description of cosmological perturbations from inflation.

The de Sitter entropy bound (\ref{eqn:NobsSend}) holds in many models of inflation that satisfy the null energy condition. However, the bound may be violated in models of inflation that violate the null energy condition, as well as those of false vacuum eternal inflation without violation of the null energy condition~\cite{ArkaniHamed:2007ky}. It is therefore important to identify the regime of validity of the de Sitter entropy bound conjecture.

\section{Implications to ghost inflation}
\label{sec:ghostinflation}

In this section let us investigate the implications of the de Sitter entropy bound conjecture to ghost inflation, which is inflation driven by ghost condensation, the simplest implementation of analogue of the Higgs mechanism in gravity. It is the simplest in the sense that the number of Nambu-Goldstone (NG) boson is only one. Before studying the de Sitter entropy bound conjecture in the context of ghost inflation, we briefly review ghost condensation and then ghost inflation.

\subsection{Ghost condensation}

In Higgs mechanism, the order parameter of spontaneous symmetry breaking is the vacuum expectation value (vev) of a scalar field, say $\Phi$, charged under the gauge group of interest. It is therefore supposed that a non-vanishing vev is developed due to a non-trivial potential such as a double-well potential. At the origin $\langle\Phi\rangle=0$ fluctuation is unstable due to the existence of a tachyon, but away from the origin $\langle\Phi\rangle\ne 0$ there are stable minima where $V'=0$ and $V''>0$. The system is thus stabilized at those stable minima and, as a result, the gauge symmetry is broken spontaneously, leading to modification of the force law in the infrared.

In ghost condensation, on the other hand, the order parameter of spontaneous symmetry breaking is the vev of derivative of a scalar field, $\langle\partial_{\mu}\phi\rangle$. It is thus supposed that the system enjoys the shift symmetry, i.e. the action of the system is invariant under a constant shift of $\phi$, $\phi\to \phi+const.$, at least in a range of field space. Let us then consider terms depending on the first derivative of $\phi$ through $X=-g^{\mu\nu}\partial_{\mu}\phi\partial_{\nu}\phi$, where $g^{\mu\nu}$ is the inverse of the spacetime metric $g_{\mu\nu}$. In the same spirit as in the tachyon instability for the Higgs mechanism, let us suppose that the Lagrangian of $\phi$ starts with $L_{\phi}=-M^4 X+\cdots$ so that the origin $\langle\partial_{\mu}\phi\rangle=0$ is unstable due to a ghost. Here $M$ is some mass scale introduced to make $X$ and $\dot{\phi}$ dimensionless. By adding nonlinear terms as $L_{\phi}=M^4[-X+X^2+\cdots]$, can we stabilize the system away from the origin $\langle\partial_{\mu}\phi\rangle\ne 0$? The answer turns out to be yes. However, since terms nonlinear in $X$ become as important as the linear term, we should consider all possible terms consistent with the symmetry as 
\begin{equation}
 L_{\phi} = \sqrt{-g}\left[P(X) + Q(X)(\Box\phi)^2 + \cdots\,\right],
\end{equation}
where we have assumed the symmetry under the field reflection $\phi\to -\phi$ as well as the shift symmetry that we have already invoked. For example, for $L_{\phi}=\sqrt{-g}P(X)$ in a flat FLRW spacetime, the equation of motion for $\phi=\phi(t)$ is $\partial_t(a^3P'\dot{\phi})=0$, meaning that either $P'\to 0$ or $\dot{\phi}\to 0$ as the universe expands ($a\to\infty$). The latter case is not allowed if the system includes a ghost at the origin $\langle\partial_{\mu}\phi\rangle=0$. The system is thus driven to the attractor with $P'=0$, meaning that $\langle\partial_{\mu}\phi\rangle\ne 0$, provided that the function $P'(X)$ has a non-zero root $X=X_0$. The stress-energy tensor for this simple system is $T_{\mu\nu}=Pg_{\mu\nu}+P'\partial_{\mu}\phi\partial_{\nu}\phi\to P(X_0)g_{\mu\nu}$, agreeing with that of a cosmological constant $-P(X_0)$. Therefore, the system at the attractor coupled with gravity allows for a Minkowski or de Sitter solution, depending on whether $P(X_0)=0$ or $P(X_0)<0$. Similar conclusion can be easily drawn in the presence of other terms such as $Q(X)(\Box\phi)^2$. One might worry about the existence of terms with higher time derivatives, but we shall later see by a proper analysis of scaling dimensions that they are irrelevant at low energies so that we can construct a consistent low-energy effective field theory describing fluctuations of the system around the attractor.

Let us then assume that a background is characterized by $\langle\partial_{\mu}\phi\rangle\ne 0$ and timelike and that the background metric is maximally symmetric, either Minkowski or de Sitter. We call this background ghost condensation. One can choose a gauge so that 
\begin{equation}
 \phi=t\,,\quad (\mbox{unitary gauge})\,.
\end{equation}
The residual symmetry is then the spatial diffeomorphism invariance, i.e. the symmetry under time-dependent spatial coordinate transformations 
\begin{equation}
 \vec{x} \to \vec{x}'(t,\vec{x})\,.
\end{equation}
We should write down most general action invariant under this residual symmetry. Starting with Minkowski background for simplicity (though extension to de Sitter background is straightforward), $g_{\mu\nu}=\eta_{\mu\nu}+h_{\mu\nu}$, the transformation under the residual symmetry transformation at the linearized level, $\vec{x}\to\vec{x}+\vec{\xi}(t,\vec{x})$, is
\begin{equation}
 h_{00}\to h_{00}\,,\quad 
  h_{0i}\to h_{0i}+\partial_0\xi_i\,, \quad
  h_{ij}\to h_{ij}+\partial_i\xi_j+\partial_j\xi_i\,.
\end{equation}
Since we assumed that Minkowski is a background satisfying the equations of motion, the action should start at the quadratic order. While $h_{00}^2$ is invariant under the residual symmetry transformation, neither $\delta^{ij}h_{0i}h_{0j}$ nor $\delta^{ik}\delta^{jl}h_{ij}h_{kl}$ is. We thus need to form a linear combination of the ($0i$) and ($ij$) components that is invariant under the residual symmetry transformation, and then square it. The invariant combination is 
\begin{equation}
 K_{ij} = \frac{1}{2}(\partial_0h_{ij}-\partial_ih_{0j}-\partial_jh_{0i})\,,
\end{equation}
and corresponds to the extrinsic curvature of constant-time hypersurfaces at the linearized level. We thus have a general Lagrangian in the unitary gauge, starting at the quadratic level as
\begin{equation}
 L_{\rm eff} = L_{\rm EH} 
  + M^4
  \left[\frac{1}{8}(h_{00})^2
   -\frac{\alpha_1}{2M^2}K^2
   -\frac{\alpha_2}{2M^2}K^{ij}K_{ij} + \cdots \right]\,,
\end{equation}
where $K=\delta^{ij}K_{ij}$ and $K^{ij}=\delta^{ik}\delta^{jl}K_{kl}$. Having constructed the general action in the unitary gauge, it is easy to obtain the action in the general gauge by simply introducing a Nambu-Goldstone (NG) boson as $t\to t+\pi$. Concretely, by replacing $h_{00}$ and $K_{ij}$ with $h_{00}-2\dot{\pi}$ and $K_{ij}+\partial_i\partial_j\pi$, respectively, we obtain
\begin{equation}
 L_{\rm eff} = L_{\rm EH} 
  + M^4
  \left[\frac{1}{2}\left(\dot{\pi}-\frac{1}{2}h_{00}\right)^2
   -\frac{\alpha_1}{2M^2}(\vec{\nabla}^2\pi+K)^2
   -\frac{\alpha_2}{2M^2}(\vec{\nabla}^i\vec{\nabla}^j\pi+K^{ij})
   (\vec{\nabla}_i\vec{\nabla}_j\pi+K_{ij}) + \cdots \right]\,.
\end{equation}

In the effective theory there are of course higher dimensional operators such as $\dot{\pi}(\nabla\pi)^2/M^2$. We now show that they are irrelevant. For this purpose we need to do proper analysis of scaling dimensions. In the decoupling limit, one can ignore the mixing between the NG boson and the metric perturbation and thus the dynamics of $\pi$ is well described by the effective action of the form
\begin{equation}
 I_{\pi} = \frac{M^4}{2}\int dtd^3\vec{x}
  \left[\dot{\pi}^2
   -\frac{\alpha}{M^2}(\vec{\nabla}^2\pi)^2 + \cdots \right]\,, 
\end{equation}
where $\alpha=\alpha_1+\alpha_2$. It is obvious that the scaling dimension of $\pi$ is not $1$ but $1/4$ as the action is invariant under the scaling $dt\to r^{-1}dt$, $d\vec{x}\to r^{-1/2}d\vec{x}$, $\pi\to r^{1/4}\pi$. It is then straightforward to calculate the scaling dimension of nonlinear operators that can enter the effective action. For example, the leading nonlinear operator is $\int dtd^3\vec{x}\dot{\pi}(\vec{\nabla}\pi)^2/M^2$ and has the scaling dimension $1/4$, which is positive. One can easily see that all other possible nonlinear operators have positive scaling dimensions, meaning that they are irrelevant at low energy (with energies and momenta $\ll M$). Those operators that exist but are irrelevant include terms with higher time derivatives, which lead to superficial new degrees of freedom. Those superficial new degrees of freedom have frequencies higher than $M$ and thus are not present in the regime of validity of the low energy effective field theory, provided that we consider $M$ as the cutoff scale. We have thus constructed a good low-energy effective field theory around the ghost condensation.

The scale $M$ sets the order parameter of the spontaneous Lorentz violation and thus general relativity is recovered in the limit $M\to 0$. We thus expect to have upper bounds on $M$ from observations and experiments. The strongest phenomenological upper bound,
\begin{equation}
 M \lesssim 100\,{\rm GeV}\,, \label{eqn:M<100GeV}
\end{equation}
comes from the stochastic scattering of cosmic microwave background (CMB) photons by lumps of positive and negative energies created due to nonlinear dynamics~\cite{ArkaniHamed:2005gu}. If we assume that the excitation of ghost condensate is responsible for all dark matter in the universe then one can obtain a (model-dependent) lower bound on $M$ as~\cite{Furukawa:2010gr}
\begin{equation}
 M \gtrsim10\,{\rm eV}\,, \quad (\mbox{model-dependent})\,.
\end{equation}

\subsection{Ghost inflation}

Ghost inflation~\cite{ArkaniHamed:2003uz} is a model of inflation driven by ghost condensation. The model is of hybrid type in the sense that the system has not only $\phi$, that is responsible for the ghost condensation, but also another field $\chi$, that plays the role of the waterfall field in hybrid inflation models. For example, one can suppose that the shift symmetry for $\phi$ is locally broken in the range $\phi_c-\Delta<\phi<\phi_c+\Delta$ so that the mass squared $m_{\chi}^2$ for $\chi$ around $\langle\chi\rangle=0$ is a positive constant for $\phi<\phi_c-\Delta$ and a negative constant for $\phi>\phi_c+\Delta$, where we have set $\dot{\phi}>0$ without loss of generality (because of the reflection symmetry in the field space). For $\phi<\phi_c-\Delta$, the background is essentially de Sitter with $\langle\chi\rangle=0$, generating scale-invariant perturbations. A soft breaking of the shift symmetry introduces deviation from the exact scale-invariance~\cite{Senatore:2004rj}. On the other hand, for $\phi>\phi_c+\Delta$, the waterfall field $\chi$ develops a non-vanishing vev, $\langle\chi\rangle\ne 0$, ending the inflation.

Since the scaling dimension of $\pi=\delta\phi$ is $1/4$, the amplitude of quantum fluctuations during ghost inflation is proportional to $(H/M)^{1/4}$. Since $\phi$ was normalized by a power of $M$ so that $\dot{\phi}=O(1)$ is dimensionless, $M\pi$ is dimensionless. Thus the amplitude of quantum fluctuation is expected to be of order $M^{-1}(H/M)^{1/4}$. Therefore, the amplitude of density perturbation is easily estimated as
\begin{equation}
 \frac{\delta\rho}{\rho} \sim \frac{H\pi}{\dot{\phi}}
  \sim \left(\frac{H}{M}\right)^{5/4}\,.
\end{equation}
This estimate agrees with the explicit calculation of the power spectrum. By requiring that this be responsible for the observed amplitude of cosmic microwave background anisotropies, $\delta\rho/\rho\sim 10^{-5}$, one thus obtains
\begin{equation}
 \frac{H}{M} \sim 10^{-4}\,.
\end{equation}
Therefore, the background curvature $\sim H^2$ is much lower than the cutoff scale $M^2$, meaning that ghost inflation is well within the regime of validity of the low-energy effective field theory. Because of the phenomenological upper bound on $M$ shown in (\ref{eqn:M<100GeV}), the Hubble expansion rate during ghost inflation is rather low ($H\lesssim10\,{\rm MeV}$), meaning that the amplitude of tensor mode is too low to be observed. While the Hubble expansion rate is low, the energy density $\rho=3M_{\rm Pl}^2H^2$ does not have to be low and thus the reheating temperature can be well above the ${\rm TeV}$ scale (up to $10^8\,{\rm GeV}$ or so).

One can calculate not only the power spectrum but also the bispectrum~\cite{ArkaniHamed:2003uz} and trispectrum~\cite{Izumi:2010wm}. Because of the higher derivative nature of the nonlinear interactions, the non-Gaussianity generated by ghost inflation is of the equilateral- and orthogonal-types, while local-type non-Gaussianity is rather small. In particular, the strongest bound on nonlinear interactions comes from the equilateral-type bispectrum and the prediction for the corresponding nonlinear parameter is
\begin{equation}
 f_{\rm NL}^{\rm equil} \simeq 82 \beta\alpha^{-4/5}\,,
\end{equation}
where $\beta$ is the coefficient of the leading nonlinear operator $\dot{\pi}(\vec{\nabla}\pi)^2/M^2$. The Planck 2015 (temperature \& polarization) constraint~\cite{Ade:2015ava} is
\begin{equation}
 f_{\rm NL}^{\rm equil} = -4 \pm 43\,, \quad (68\% \mbox{ CL statistical})\,
\end{equation}
leading to the constraint on the coefficient of the nonlinear operator as
\begin{equation}
 -0.6 \lesssim \beta\alpha^{-4/5} \lesssim 0.5\,.
\end{equation}

\subsection{Lower bound on $\Omega_{\Lambda}$}
\label{subsec:OmegaLambda}

Having briefly reviewed ghost condensation and ghost inflation, we are now ready to consider implications of the de Sitter entropy bound on ghost inflation. While the de Sitter entropy is still given by $S=\pi/(G_{\rm N}H^2)$, the total number of e-foldings $N_{\rm tot}$ is arbitrary. Hence the de Sitter entropy bound (\ref{eqn:NobsSend}) is threaten to be violated. Ref.~\cite{ArkaniHamed:2007ky} then argued that ghost inflation spoils the thermodynamic interpretation of the de Sitter entropy and thus is incompatible with basic gravitational principles. In this subsection, on the contrary, we shall show that the de Sitter entropy bound (\ref{eqn:NobsSend}) is indeed satisfied in our universe very well when we assume that ghost inflation is responsible for the observed CMB anisotropies. What is important here is that the left hand side of the de Sitter entropy bound (\ref{eqn:NobsSend}) is not the total number of e-foldings $N_{\rm tot}$ but the observable number of e-foldings $N_{\rm obs}$ and that superhorizon modes never reenter the horizon in the late time universe dominated by dark energy. 

For simplicity we suppose that after the end of ghost inflation at $t=t_{\rm end}$, our universe experienced a matter dominated (MD) epoch due to the oscillation of the waterfall field $\chi$, followed by a reheating at $t=t_{\rm reh}$ and then a radiation dominated (RD) epoch until the radiation-matter equality at $t=t_{\rm eq}$. After that, the universe is supposed to evolve according to the $\Lambda$CDM cosmology (but with a priori arbitrary parameter values). Just for simplicity we also assume that the late-time acceleration ($\ddot{a}>0$) starts well after the matter-radiation equality as in our universe. We shall then seek a lower bound on $\Omega_{\Lambda}\equiv\Lambda/(3H_0^2)$ by requiring that the de Sitter entropy bound for ghost inflation be satisfied, where $\Lambda$ is the effective cosmological constant (that is the sum of the genuine cosmological constant and the contribution from the ghost condensation) and $H_0$ is the present value of the Hubble expansion rate.

\begin{figure}
 \begin{center}
 \includegraphics[width=0.5\textwidth,clip]{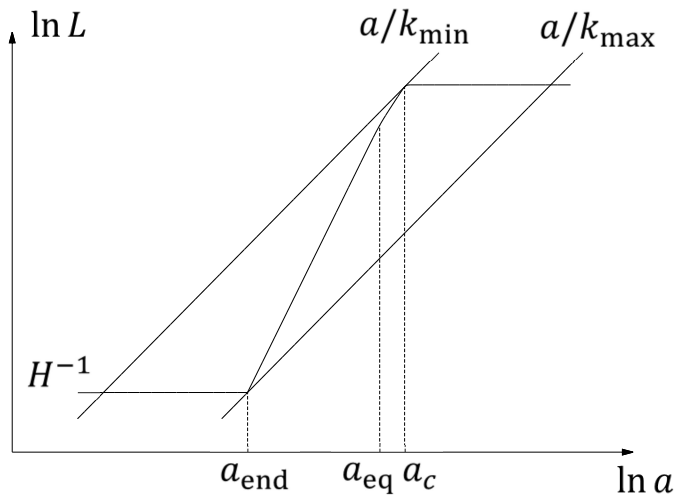}
 \end{center}
 \caption{\label{fig:horizonscale}
Evolutions of the Hubble horizon scale and comoving scales. There are the maximum and minimum comoving momenta, $k_\mathrm{max}$ and $k_\mathrm{min}$, that exit the horizon during ghost inflation and then reenter the horizon. On the horizontal axis, $a_\mathrm{end}$, $a_\mathrm{eq}$ and $a_c$ stand for their logarithms.
}
\end{figure}

\begin{figure}
 \begin{center}
 \includegraphics[width=0.5\textwidth,clip]{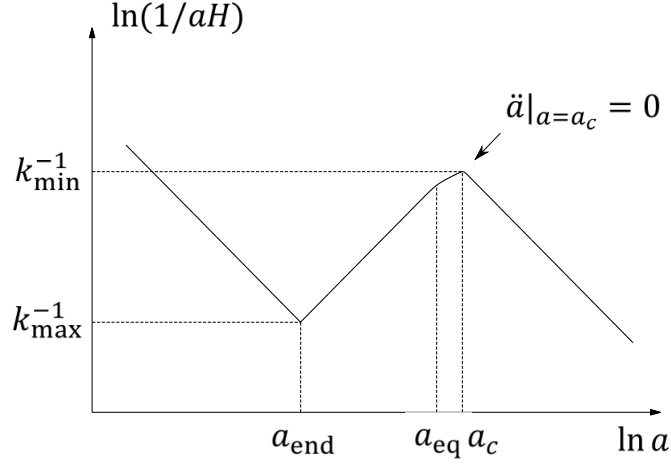}
 \end{center}
 \caption{\label{fig:1/adot}
 Evolution of the comoving Hubble horizon scale. The comoving scale corresponding to $k_\mathrm{min}$ reenters the horizon at $a=a_c$, where $a_c$($\gg a_\mathrm{eq}$) is defined by $\ddot{a}|_{a=a_c}=0$. On the horizontal axis, $a_\mathrm{end}$, $a_\mathrm{eq}$, and $a_c$ stand for logarithms of them.
 }
\end{figure}

The overall evolutions of the Hubble horizon scale and comoving scales are plotted in Fig.~\ref{fig:horizonscale}. Also, since the comoving momentum $k$ exits/reenters the horizon when $k=aH$, we show the behavior of $1/(aH)=1/\dot{a}$ in Fig.~\ref{fig:1/adot}. In this situation under consideration, there are the maximum and minimum comoving momenta, $k_{\rm max}$ and $k_{\rm min}$, that exit the horizon during ghost inflation and then reenter the horizon in the late time universe. The comoving scale corresponding to $k_{\rm max}$ exits and reenters the horizon just at the end of ghost inflation. The comoving scale corresponding to $k_{\rm min}$ reenters the horizon at $t=t_c$, where $t_c$ ($\gg t_{\rm eq}$) is defined by $\ddot{a}(t=t_c)=0$. 
%
%

Let us now estimate $a_c\equiv a(t_c)$ and $k_{\rm min}$. In the late time epoch with $a\gg a_{\rm eq}$ ($\equiv a(t=t_{\rm eq})$), we can safely assume that the universe is dominated by the matter and the cosmological constant so that
\begin{equation}
 3M_{\rm Pl}^2H^2 = \rho_{\rm m}^{\rm eq}\left(\frac{a_{\rm eq}}{a}\right)^3 + \rho_{\Lambda}\,, \quad
 6M_{\rm Pl}^2\frac{\ddot{a}}{a} = -\rho_{\rm m}^{\rm eq}\left(\frac{a_{\rm eq}}{a}\right)^3 + 2\rho_{\Lambda}\,, 
\end{equation}
where $\rho_{\rm m}^{\rm eq}$ is the matter energy density at $t=t_{\rm eq}$ and $\rho_{\Lambda}=M_{\rm Pl}^2\Lambda=3M_{\rm Pl}^2H_0^2\Omega_{\Lambda}$ is the energy density corresponding to the cosmological constant $\Lambda$, $H_0$ is the Hubble expansion rate at the present time and $\Omega_{\Lambda}$ is the density parameter for $\Lambda$. Setting $\ddot{a}(t=t_c)=0$, we then obtain
\begin{equation}
 \left(\frac{a_c}{a_{\rm eq}}\right)^3 = \frac{\rho_{\rm m}^{\rm eq}}{2\rho_{\Lambda}} \sim \frac{g_{*{\rm eq}}T_{\rm eq}^4}{M_{\rm Pl}^2H_0^2\Omega_{\Lambda}}\,, \quad
  \left(\frac{H_c}{H_0}\right)^2 = 3\Omega_{\Lambda}\,, \quad
  k_{\rm min} = H_ca_c\,,
\end{equation}
where $g_{*{\rm eq}}$ is the total number of effectively massless degrees of freedom for energy density at $t=t_{\rm eq}$.

Next let us estimate $a_{\rm end}$ and $k_{\rm max}$. Since we assumed that the universe between the end of inflation and the reheating is dominated by the oscillation of the waterfall field, we have 
\begin{equation}
\frac{a_{\rm end}}{a_{\rm reh}} \sim 
 \left(\frac{\rho_{\rm reh}}{\rho_{\rm inf}}\right)^{1/3}\,,
\end{equation}
where $\rho_{\rm inf}$ and $\rho_{\rm reh}$ are the energy density at the end of inflation and at the reheating. Assuming for simplicity that there is no significant entropy production between the reheating and the matter-radiation equality, we have 
\begin{equation}
 \frac{a_{\rm reh}}{a_{\rm eq}} 
  \sim \left(\frac{s_{\rm eq}}{s_{\rm reh}}\right)^{1/3} 
  \sim \left(\frac{g_{*s,{\rm eq}}}{g_{*s,{\rm reh}}}\right)^{1/3}
  \frac{T_{\rm eq}}{T_{\rm reh}}\,,
\end{equation}
where $s_{\rm reh}$ and $s_{\rm eq}$ are the entropy density at $t=t_{\rm reh}$ and $t=t_{\rm eq}$, respectively, $g_{*s,{\rm reh}}$ and $g_{*s,{\rm eq}}$ are the total number of effectively massless degrees of freedom for entropy density at $t=t_{\rm reh}$ and $t=t_{\rm eq}$, respectively, and $T_{\rm reh}$ and $T_{\rm eq}$ are the photon temperature at $t=t_{\rm reh}$ and $t=t_{\rm eq}$, respectively. In the last expression, $T_{\rm reh}$ can be rewritten as
\begin{equation}
 T_{\rm reh} \sim 
  \left(\frac{\rho_{\rm reh}}{\rho_{\rm inf}}\right)^{1/4}
  \left(\frac{3M_{\rm Pl}^2H_{\rm inf}^2}{g_{*{\rm reh}}}\right)^{1/4}\,.
\end{equation}
We then obtain
\begin{equation}
 \frac{a_{\rm end}}{a_{\rm eq}}
  \sim
  \left(\frac{\rho_{\rm reh}}{\rho_{\rm inf}}\right)^{1/3}
  \left(\frac{g_{*s,{\rm eq}}}{g_{*s,{\rm reh}}}\right)^{1/3}
  T_{\rm eq}
  \left(\frac{\rho_{\rm reh}}{\rho_{\rm inf}}\right)^{-1/4}
  \left(\frac{3M_{\rm Pl}^2H_{\rm inf}^2}{g_{*{\rm reh}}}\right)^{-1/4}\,,
  \quad
  k_{\rm max} = H_{\rm inf}a_{\rm end}\,.
\end{equation}

The observable number of e-foldings is then~\footnote{For example, if we observe the inflationary modes through the cosmic microwave background anisotropies then the observable number of e-foldings should be given by the logarithm of the ratio of the comoving size of the Hubble horizon (or the apparent horizon) at the end of inflation to the comoving size of the last scattering surface. This agrees with $N_{\rm obs}$ in the main text up to an $O(1)$ factor.}
\begin{equation}
\label{Nobs}
 \exp (N_{\rm obs}) \sim \frac{k_{\rm max}}{k_{\rm min}}
  \sim 
  \left(\frac{\rho_{\rm reh}}{\rho_{\rm inf}}\right)^{1/12}
  \left(\frac{g_{*s,{\rm eq}}}{g_{*s,{\rm reh}}\ g_{*{\rm eq}}}\right)^{1/3}
  (g_{*{\rm reh}})^{1/4}
  \left(\frac{M_{\rm Pl}H_{\rm inf}^3}{\Omega_{\Lambda}H_0^2T_{\rm eq}^2}\right)^{1/6}\,.
\end{equation}
Demanding that $N_{\rm obs}\lesssim S_{\rm end}=8\pi^2M_{\rm Pl}^2/H_{\rm inf}^2$, one obtains a lower bound on $\Omega_{\Lambda}$ as
\begin{equation}
 \Omega_{\Lambda} \gtrsim
  \left(\frac{\rho_{\rm reh}}{\rho_{\rm inf}}\right)^{1/2}
  \left(\frac{g_{*s,{\rm eq}}}{g_{*s,{\rm reh}}\ g_{*{\rm eq}}}\right)^2
  (g_{*{\rm reh}})^{3/2}
  \left(\frac{M_{\rm Pl}H_{\rm inf}^3}{H_0^2T_{\rm eq}^2}\right)
  \times \exp(-6S_{\rm end})\,.
\end{equation}
Since 
\begin{equation}
 6S_{\rm end} \sim \frac{48\pi^2M_{\rm Pl}^2}{(10^{-4}\, M)^2} \sim 
  10^{42}\left(\frac{M}{100\,{\rm GeV}}\right)^{-2}\,,\quad M \lesssim 100\,{\rm GeV}\,,
\end{equation}
one can safely neglect the prefactor and the lower bound then becomes
\begin{equation}
 \Omega_{\Lambda} \gtrsim \exp\left[-10^{42}\left(\frac{M}{100\,{\rm GeV}}\right)^{-2}\right]\,. \label{eqn:lowerbound-OmegaLambda}
\end{equation}
Since $M \lesssim 100\,{\rm GeV}$, the right hand side is extremely tiny and thus the bound is satisfied in our universe very well.

\section{Cosmological Page time}
\label{sec:pagetime}

In the previous section we have shown that ghost inflation indeed satisfies the de Sitter entropy bound if it happened in the early epoch of our universe. However, if the observed effective cosmological constant $\Lambda$ is not eternal but decays to zero or an extremely small positive value in the future then the entropy bound may in principle be violated in the distant future (but not yet now). In order to understand what would happen in such a universe in the distant future, let us first remind ourselves about the black hole evaporation, following the argument made by Page~\cite{Page:1993wv}.

\subsection{Black hole evaporation and information}
\label{subsec:information}

\begin{figure}
 \begin{center}
 \includegraphics[width=0.5\textwidth,clip]{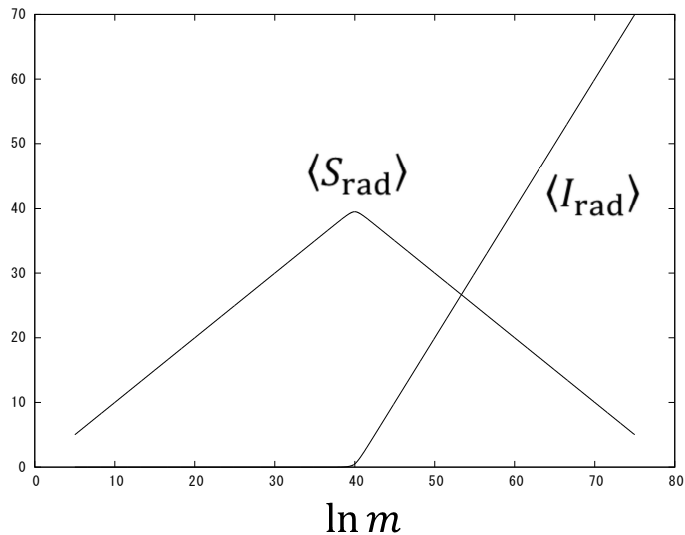}
 \end{center}
 \caption{\label{fig:average}
 The average entanglement entropy $\langle S_\mathrm{rad}\rangle$ and the average information $\langle I_\mathrm{rad}\rangle$ of a radiation subsystem versus its thermodynamic entropy $\ln m$. The dimension of the entire Hilbert space is fixed to be $mn=e^{80}$.
 }
\end{figure}

Let us consider a Hilbert space of the form $\mathcal{H}=\mathcal{H}_{\rm rad}\otimes\mathcal{H}_{\rm bh}$ with $\dim\mathcal{H}_{\rm rad}=m$ and $\dim\mathcal{H}_{\rm bh}=n$. For a pure state $|\phi\rangle$ ($\in{\mathcal H}$), one can define the reduced density matrices $\rho_{\rm rad}\equiv \mathrm{Tr}_{\rm bh}\rho$ and $\rho_{\rm bh}\equiv \mathrm{Tr}_{\rm rad}\rho$, where $\mathrm{Tr}_{\rm bh}$ and $\mathrm{Tr}_{\rm rad}$ represent trace over $\mathcal{H}_{\rm bh}$ and $\mathcal{H}_{\rm rad}$, respectively, and $\rho=|\phi\rangle\langle\phi|$. The entanglement entropy is then defined as $S_{\rm rad}=-\mathrm{Tr}_{\rm rad}\rho_{\rm rad}\ln\rho_{\rm rad}$ or $S_{\rm bh}=-\mathrm{Tr}_{\rm bh}\rho_{\rm bh}\ln\rho_{\rm bh}$, and they agree, $S_{\rm rad}=S_{\rm bh}$. Information in $\rho_{\rm rad}$ and $\rho_{\rm bh}$ are, respectively, defined as $I_{\rm rad}=\ln m-S_{\rm rad}$ and $I_{\rm bh}=\ln n-S_{\rm bh}$. For $1\ll m\leq n$, Page~\cite{Page:1993df} estimated the average entanglement entropy $\langle S_{\rm rad}\rangle$ and the average information $\langle I_{\rm rad}\rangle$ as
\begin{equation}
 \langle I_{\rm rad}\rangle = \ln m - \langle S_{\rm rad}\rangle
  \simeq \frac{m}{2n}\,, \quad (1\ll m\leq n)\,,
\end{equation}
where the averaging is defined using the unitarily invariant Haar measure on $\mathcal{H}$. By using the fact that $S_{\rm rad}=S_{\rm bh}$, it is then obvious that
\begin{equation}
 \langle S_{\rm rad}\rangle
  \simeq \ln n - \frac{n}{2m}\,, \quad (1\ll n\leq m)\,,
\end{equation}
and that 
\begin{equation}
 \langle I_{\rm rad}\rangle
  \simeq \ln\frac{m}{n} + \frac{n}{2m}\,, \quad (1\ll n\leq m)\,.
\end{equation}
Fig.~\ref{fig:average} shows plots of $\langle S_{\rm rad}\rangle$ and $\langle I_{\rm rad}\rangle$. 
%

From this result one can extract important implication on an evaporating black hole by setting $\ln n$ to be the black hole entropy, $4\pi M_{\rm bh}^2/M_{\rm Pl}^2$~\cite{Page:1993wv}. For the early stage ($1\ll m\ll n$) of an evaporating semi-classical ($M_{\rm bh}\gg M_{\rm Pl}$) black hole, one obtains $\langle I_{\rm rad}\rangle\simeq m/(2n)\sim m\exp(-4\pi M_{\rm bh}^2/M_{\rm Pl}^2)$. We thus have 
\begin{equation}
 r_{\rm bh}\frac{d}{dt}\langle I_{\rm rad}\rangle \sim \exp\left(-\frac{4\pi}{y^2}\right)\,, \quad r_{\rm bh} = \frac{2M_{\rm bh}}{M_{\rm Pl}^2}, \quad y\equiv \frac{M_{\rm Pl}}{M_{\rm bh}}\ll 1\,.
\end{equation}
This is not analytic in $y$ and thus one can never recover information by perturbation in $y$. That is, if one observes only a small subsystem of a large system at one time then the amount of information to be recovered is tiny. The total information is instead encoded in the entire system. Actually, from Fig.~\ref{fig:average} one can clearly see that the first bit of information comes out from a black hole at the so called Page time, the time when the black hole loses about half of its entropy. After the Page time, information comes out.

For the early stage of black hole evaporation, i.e. sufficiently earlier than the Page time, the above result based on the average information perfectly matches with that based on effective field theory: both fail to reconstruct information. On the other hand, starting at the Page time, the result based on the average information predicts recovery of information. A problem with effective field theory is that it continues to fail to reconstruct information, even after the Page time. This means that the prediction of effective field theory starts to deviate from the correct result by $O(1)$ at the Page time.

Needless to say, the expected $O(1)$ deviation from effective field theory prediction neither implies the end of the world at the Page time nor invalidates the effective field theory description of the early stage of black hole evaporation.

\subsection{Definition of cosmological Page time}

From the considerations in subsection~\ref{subsec:information} it is concluded that effective field theory should break down when it relies on about half of the number of states behind the horizon. Extrapolating this conclusion to the de Sitter horizon, we find it natural to define the cosmological Page time $t_{\rm Page}$ as the time (after inflation) when the number of inflationary modes that reenter the horizon becomes as large as $\exp(S_{\rm end})$, where $S_{\rm end}$ is the de Sitter entropy at the end of inflation. In other words, we define the cosmological Page time $t_{\rm Page}$ as the time when the de Sitter entropy bound (\ref{eqn:NobsSend}) starts to be violated.

\subsection{Cosmological Page time after ghost inflation}

As we have seen in subsection \ref{subsec:OmegaLambda}, if ghost inflation happened in the early epoch of our universe then the de Sitter entropy bound is satisfied. This means that $t_{\rm Page}=\infty$ in our universe, provided that the effective cosmological constant $\Lambda$ that we observe today is eternal. 

On the other hand, if the observed $\Lambda$ is not eternal but decays to an extremely small value or zero in the future then there is a possibility that the de Sitter entropy bound (\ref{eqn:NobsSend}) may be violated in the distant future.

\begin{figure}
 \begin{center}
 \includegraphics[width=0.5\textwidth,clip]{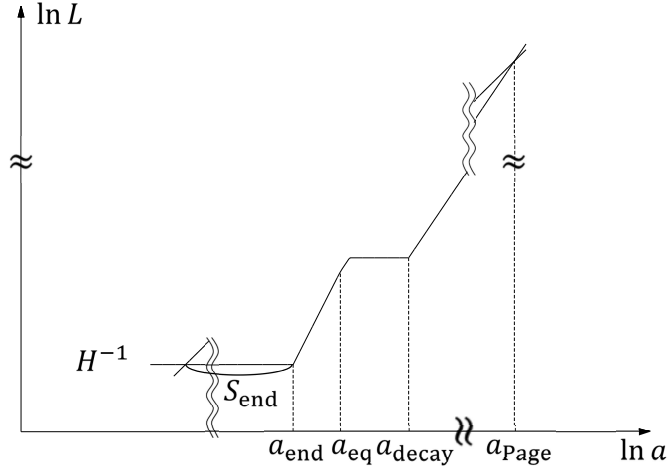}
 \end{center}
 \caption{\label{fig:decaying}
Evolutions of the Hubble horizon scale and a comoving scale in the case where the cosmological constant decays at $a=a_\mathrm{decay}$ and the universe becomes matter-dominated. The number of inflationary modes that reenter the horizon becomes as large as $\exp(S_{\rm end})$ at $a=a_\mathrm{Page}$. On the horizontal axis, $a_\mathrm{end}$, $a_\mathrm{eq}$, $a_\mathrm{decay}$ and $a_\mathrm{Page}$ stand for their logarithms. For $a>a_{\rm Page}$, the Hubble horizon scale exceeds the comoving scale corresponding to $k=a_{\rm Page}H_{\rm Page}=a_{\rm end}H_{\rm inf}e^{-S_{\rm end}}$ and thus observers may see $O(1)$ deviation from the EFT prediction. 
}
\end{figure}

For example, imagine that our current dark energy dominated universe is metastable such that it completely decays after a while at $t=t_{\rm decay}$ to matter. One can think of this model as a slow rolling scalar field which later enters an oscillatory phase. Alternatively, one can introduce yet another waterfall field whose squared mass changes from a positive constant to a negative constant at a certain value of the scalar field $\phi$ responsible for ghost condensation, similarly to the case of ghost inflation. Provided that the oscillation of the waterfall field does not decay to radiation (but may decay to heavy fields), the universe after the phase transition becomes matter dominated. In these cases, the overall evolutions of the Hubble horizon scale and the comoving scale corresponding to $k=a_{\rm Page}H_{\rm Page}$ are shown in Fig.~\ref{fig:decaying}, where $a_{\rm Page}$ and $H_{\rm Page}$ are the values of $a$ and $H$ at $t=t_{\rm Page}$. Since $aH\propto a^{-1/2}$ in the matter dominated epoch, we obtain
\begin{equation}
 a_{\rm Page}H_{\rm Page} = \left(\frac{a_{\rm decay}}{a_{\rm Page}}\right)^{1/2}a_{\rm decay}H_{\Lambda}\,,
\end{equation}
where $a_{\rm decay}$ is the value of $a$ at $t=t_{\rm decay}$, and $H_{\Lambda}=\sqrt{\rho_{\Lambda}/(3M_{\rm Pl}^2)}=\sqrt{\Lambda/3}$. Combining this with the estimate of $a_{\rm end}$ in subsection \ref{subsec:OmegaLambda}, we obtain
\begin{equation}
 \frac{a_{\rm end}H_{\rm inf}}{a_{\rm Page}H_{\rm Page}}
  \sim 
  \left(\frac{\rho_{\rm reh}}{\rho_{\rm inf}}\right)^{1/12}
  \left(\frac{g_{*s,{\rm eq}}}{g_{*s,{\rm reh}}}\right)^{1/3}(g_{*{\rm reh}})^{1/4}
  \left(\frac{T_{\rm eq}^4H_{\rm inf}^2}{3M_{\rm Pl}^2H_{\Lambda}^4}\right)^{1/4}
  \frac{a_{\rm eq}}{a_{\rm decay}}\left(\frac{a_{\rm Page}}{a_{\rm decay}}\right)^{1/2}. 
\end{equation}
By equating this with $e^{S_{\rm end}}$ and using $H_{\rm inf}\sim 10^{-4}M$, we obtain
\begin{equation}
 \frac{a_{\rm Page}}{a_{\rm decay}} \sim 
  \left(\frac{\rho_{\rm inf}}{\rho_{\rm reh}}\right)^{1/6}
  \left(\frac{g_{*s,{\rm reh}}}{g_{*s,{\rm eq}}}\right)^{2/3}(g_{*{\rm reh}})^{-1/2}
  \left[\frac{3M_{\rm Pl}^2H_{\Lambda}^4}{(10^{-2}\,{\rm GeV})^2T_{\rm eq}^4}\right]^{1/2}
  \left(\frac{M}{100\,{\rm GeV}}\right)^{-1}
  \left(\frac{a_{\rm decay}}{a_{\rm eq}}\right)^2
  \exp(2S_{\rm end}).
\end{equation}
By substituting the estimate of $S_{\rm end}$ in subsection \ref{subsec:OmegaLambda}, we finally obtain
\begin{equation}
 \frac{a_{\rm Page}}{a_{\rm decay}} \sim 
  \left(\frac{M}{100\,{\rm GeV}}\right)^{-1}
  \left(\frac{a_{\rm decay}}{a_{\rm eq}}\right)^2
  \exp\left[10^{42}  \left(\frac{M}{100\,{\rm GeV}}\right)^{-2}\right].
\end{equation}
Therefore, it is in an extremely distant future when the universe reaches the Page time after the observed $\Lambda$ decays. 
%

For $t\ll t_{\rm Page}$ (including the present time), effective field theory is expected to be a good description of inflationary modes, predicting correct correlation functions among cosmological observables. At around $t\sim t_{\rm Page}$, however, observers start to see $O(1)$ deviation from the prediction of effective field theory. For example, one might observe unexpected correlations among superficially independent modes. This is because at around $t\sim t_{\rm Page}$ the number of observable modes starts to exceed the number of physical degrees of freedom accessible during inflation, and thus those superficially independent, observable modes cannot be independent of each other.

As in the case of black hole evaporation, the expected $O(1)$ deviation from effective field theory prediction neither implies the end of the world at around $t\sim t_{\rm Page}$ nor invalidates the effective field theory description of cosmological observables for $t\ll t_{\rm Page}$.

\section{Summary and discussions}
\label{sec:summary}

We have revisited the issue of de Sitter entropy in ghost inflation. As we have briefly reviewed in Sec.~\ref{sec:dSentropy}, it was conjectured in \cite{ArkaniHamed:2007ky} that the number of e-foldings that can be observed by late time observers should be bounded from above by the de Sitter entropy at the end of inflation $S_{\rm end}$. This conjecture is supported by a number of evidences. For example, the bound holds in the standard slow-roll inflation. Also, in slow-roll eternal inflation, while the total number of e-foldings can be significantly larger than the de Sitter entropy, the formation of a black hole prevents the extra e-foldings from being observed by late time observers. In Sec.~\ref{sec:ghostinflation} we have then investigated the implications of the de Sitter entropy bound conjecture to ghost inflation. In the ghost inflation (as in the slow-roll eternal inflation), the total e-folding number is arbitrary in principle, and thus there is a potential danger that the model might break the de Sitter entropy bound. This may be an interesting opportunity in the sense that the de Sitter entropy bound conjecture might provide us with a non-trivial constraint on cosmological parameters. Intriguingly, as we have shown in subsection~\ref{subsec:OmegaLambda}, the de Sitter entropy bound can be restated as a lower bound on the value of the (effective) cosmological constant at the present time. This is simply because the maximum comoving scale which had been created in the inflationary era and reentered into the Hubble horizon is constrained from above by the horizon scale of the (effective) cosmological constant at present, and it gives an upper bound on the value of the observable number of e-foldings. This means that if we adopt the ghost inflation model for the early accelerated expansion of the universe, we can obtain the lower bound on the (effective) cosmological constant at late time by the de Sitter entropy bound. Fortunately or unfortunately, the lower bound is as weak as $\Omega_{\Lambda} \gtrsim\exp(-10^{42})$. Compared with this extremely tiny lower bound, the observed value of the cosmological constant in our present universe is tremendously large. Thus the de Sitter entropy bound is always satisfied for the ghost inflation in our universe.

Moreover, even if the (effective) cosmological constant decays to essentially zero sometime in the future so that the de Sitter entropy bound is eventually violated in the further distant future, it means neither that the ghost inflation is invalid at all time from the beginning to the end as the EFT nor that we will be crushed by black holes surrounding us as in the eternal slow-roll inflation case. We have drawn this conclusion in Sec.~\ref{sec:pagetime}, by comparison with the case of an evaporating black hole. According to the Page's argument on the black hole evaporation, we can investigate the Hawking radiation with the use of the semi-classical EFT, as far as the dimension of the Hilbert space associated with radiated modes is sufficiently less than that associated with the black hole. As the evaporation proceeds and the Hilbert space associated with the radiation becomes as large as the black hole's, or the system reaches the Page time, then we for the first time become unable to use the EFT to calculate physical quantities involving modes more than ${\rm exp}(S_{\rm bh})$. This is interpreted as follows: after the Page time, the uncertainties in correlations among the radiation modes as many as the dimension of the Hilbert space of the black hole accumulates as much as giving $O(1)$ deviation from the prediction of the EFT. As a result, the EFT description is expected to break down at around the Page time. However, after the Page time, we just become unable to use the EFT to make correct predictions, and if the black hole continues to evaporate, we can still watch the black hole evaporation perhaps through observing the radiation.

As an analog of the Page's argument on the black hole evaporation, we can safely use the EFT in the ghost inflation model to calculate correlations among inflationary modes that reenter the horizon well before the cosmological Page time $t_{\rm Page}$, which is defined as the time when the observed e-folding number becomes roughly equal to the de Sitter entropy at the end of inflation. We can use the EFT in the ghost inflation well before $t_{\rm Page}$ since the uncertainties in correlations among the different Hubble patches and the inflationary modes have not yet accumulated as much as giving $O(1)$ deviation from the prediction of the EFT. Only at and after around $t_{\rm Page}$, we become unable to use the EFT for the calculation of quantities involving the Hubble patches and/or the inflationary modes more than ${\rm exp}(S_{\rm end})$ since the uncertainties in correlations accumulate as much as giving $O(1)$ deviation from the prediction of the EFT. Whilst in the eternal slow-roll inflation model we will be crushed by the black hole after $t_{\rm Page}$, we do not expect such a catastrophic event after $t_{\rm Page}$ in the ghost inflation model. Therefore, as in the case of black hole evaporation, it is in principle possible in the late time universe after ghost inflation to keep observing further inflationary modes although the EFT prediction is not necessarily useful after $t_{\rm Page}$.

Based on Page's argument reviewed in subsection \ref{subsec:information}, if one observes only a small subsystem of a large system at one time then it is expected that the EFT does not result in any contradiction: both the EFT and the entropy argument would conclude that the amount of information to be recovered is tiny. For this reason, if we merely intend to describe part of the Hawking radiation a one time, then we may still be able to use the EFT even after the Page time, as far as the dimension of the Hilbert space associated with the part of radiation that we intend to describe is sufficiently smaller than that of the Hilbert space for the rest of system. In the ghost inflation model, as in the black hole evaporation, we are not able to use the EFT to describe physical quantities involving the Hubble patches and/or the inflationary modes more than ${\rm exp}(S_{\rm end})$, for example, $n$-point correlation functions with too large $n$. On the other hand, we may safely use the EFT if we merely intend to describe physical quantities involving smaller number of Hubble patches and inflationary modes, even after $t_{\rm Page}$.

In summary, the ghost inflation always satisfies the de Sitter entropy bound if the (effective) cosmological constant of the present universe keeps its value for the future. Even if the (effective) cosmological constant decays to zero or an extremely small value sometime in the future, it is in an extremely distant future $t\sim t_{\rm Page}$ when the de Sitter entropy bound is eventually violated. At around $t_{\rm Page}$, we for the first time become unable to use the EFT to describe physical quantities involving Hubble patches and/or inflationary modes more than ${\rm exp}(S_{\rm end})$. After that, although we expect $O(1)$ deviation from the EFT prediction, we can still survive the Page time and keep observing further modes reentering the horizon. Without knowing the UV-completion for ghost inflation and general relativity it is not easy to predict what the future observer will detect on his/her sky after the Page time. Nevertheless, it is not far from intuition that he/she might see unforecasted correlations between very early and very late horizon-reentered modes and that he/she cannot describe those novel correlations by the EFT that people today know of.

\section*{\bf Acknowledgments}
This work was supported in part by Grant-in-Aid for Scientific Research No. 24540256 (S.M.) and No. 16J06266 (Y.W.). The work of S.M. and Y.W. was supported in part by World Premier International Research Center Initiative (WPI), Ministry of Education, Culture, Sports, Science and Technology (MEXT), Japan. The basic idea of the present paper was developed when S.M. was visiting Institut Astrophysique de Paris (IAP). During the visit, part of his work has been done within the Labex ILP (reference ANR-10-LABX-63) part of the Idex SUPER, and received financial state aid managed by the Agence Nationale de la Recherche, as part of the programme Investissements d’avenir under the reference ANR-11- IDEX-0004-02. He is thankful to colleagues at IAP for warm hospitality. S.J. would like to thank Yukawa Institute for Theoretical Physics (YITP) for their kind hospitality during his visit, when some parts of the project were in hand. He also wants to thank H. Firouzjahi and M. Noorbala for their useful comments. The work of R.S. was supported in part by the Natural Science Foundation of China under Grants No. 11475065. R.S. thanks people in YITP for their warm hospitality during his stay at YITP. The work of Y.W. was supported by the Program for Leading Graduate Schools and WPI, MEXT, Japan. He is grateful to YITP, where part of his work was done during his visit.


\begin{thebibliography}{99}

\bibitem{Bardeen:1973gs} 
  J.~M.~Bardeen, B.~Carter and S.~W.~Hawking,
  Commun.\ Math.\ Phys.\  {\bf 31}, 161 (1973).
  doi:10.1007/BF01645742

\bibitem{Bekenstein:1973ur} 
  J.~D.~Bekenstein,
  Phys.\ Rev.\ D {\bf 7}, 2333 (1973).
  doi:10.1103/PhysRevD.7.2333

\bibitem{Hawking:1974sw} 
  S.~W.~Hawking,
  Commun.\ Math.\ Phys.\  {\bf 43}, 199 (1975)
  [Commun.\ Math.\ Phys.\  {\bf 46}, 206 (1976)].
  doi:10.1007/BF02345020


\bibitem{ArkaniHamed:2003uy} 
  N.~Arkani-Hamed, H.~C.~Cheng, M.~A.~Luty and S.~Mukohyama,
  JHEP {\bf 0405}, 074 (2004)
  doi:10.1088/1126-6708/2004/05/074
  [hep-th/0312099].

\bibitem{Dubovsky:2006vk} 
  S.~L.~Dubovsky and S.~M.~Sibiryakov,
  Phys.\ Lett.\ B {\bf 638}, 509 (2006)
  doi:10.1016/j.physletb.2006.05.074
  [hep-th/0603158].

\bibitem{Eling:2007qd} 
  C.~Eling, B.~Z.~Foster, T.~Jacobson and A.~C.~Wall,
  Phys.\ Rev.\ D {\bf 75}, 101502 (2007)
  doi:10.1103/PhysRevD.75.101502
  [hep-th/0702124 [HEP-TH]].

\bibitem{Jacobson:2008yc} 
  T.~Jacobson and A.~C.~Wall,
  Found.\ Phys.\  {\bf 40}, 1076 (2010)
  doi:10.1007/s10701-010-9423-5
  [arXiv:0804.2720 [hep-th]].

\bibitem{Mukohyama:2009rk} 
  S.~Mukohyama,
  JHEP {\bf 0909}, 070 (2009)
  doi:10.1088/1126-6708/2009/09/070
  [arXiv:0901.3595 [hep-th]].

\bibitem{Mukohyama:2009um} 
  S.~Mukohyama,
  Open Astron.\ J.\  {\bf 3}, 30 (2010)
  doi:10.2174/1874381101003020030
  [arXiv:0908.4123 [hep-th]].

\bibitem{Israel:1976ur} 
  W.~Israel,
  Phys.\ Lett.\ A {\bf 57}, 107 (1976).
  doi:10.1016/0375-9601(76)90178-X

\bibitem{Frolov:2002va} 
  A.~V.~Frolov and L.~Kofman,
  JCAP {\bf 0305}, 009 (2003)
  doi:10.1088/1475-7516/2003/05/009
  [hep-th/0212327].

\bibitem{Goheer:2002vf} 
  N.~Goheer, M.~Kleban and L.~Susskind,
  JHEP {\bf 0307}, 056 (2003)
  doi:10.1088/1126-6708/2003/07/056
  [hep-th/0212209].

\bibitem{Kachru:2003aw} 
  S.~Kachru, R.~Kallosh, A.~D.~Linde and S.~P.~Trivedi,
  Phys.\ Rev.\ D {\bf 68}, 046005 (2003)
  doi:10.1103/PhysRevD.68.046005
  [hep-th/0301240].

\bibitem{ArkaniHamed:2007ky} 
  N.~Arkani-Hamed, S.~Dubovsky, A.~Nicolis, E.~Trincherini and G.~Villadoro,
  JHEP {\bf 0705}, 055 (2007)
  doi:10.1088/1126-6708/2007/05/055
  [arXiv:0704.1814 [hep-th]].

\bibitem{ArkaniHamed:2003uz} 
  N.~Arkani-Hamed, P.~Creminelli, S.~Mukohyama and M.~Zaldarriaga,
  JCAP {\bf 0404}, 001 (2004)
  doi:10.1088/1475-7516/2004/04/001
  [hep-th/0312100].

\bibitem{ArkaniHamed:2005gu} 
  N.~Arkani-Hamed, H.~C.~Cheng, M.~A.~Luty, S.~Mukohyama and T.~Wiseman,
  JHEP {\bf 0701}, 036 (2007)
  doi:10.1088/1126-6708/2007/01/036
  [hep-ph/0507120].

\bibitem{Furukawa:2010gr} 
  T.~Furukawa, S.~Yokoyama, K.~Ichiki, N.~Sugiyama and S.~Mukohyama,
  JCAP {\bf 1005}, 007 (2010)
  doi:10.1088/1475-7516/2010/05/007
  [arXiv:1001.4634 [astro-ph.CO]].

\bibitem{Senatore:2004rj} 
  L.~Senatore,
  Phys.\ Rev.\ D {\bf 71}, 043512 (2005)
  doi:10.1103/PhysRevD.71.043512
  [astro-ph/0406187].

\bibitem{Izumi:2010wm} 
  K.~Izumi and S.~Mukohyama,
  JCAP {\bf 1006}, 016 (2010)
  doi:10.1088/1475-7516/2010/06/016
  [arXiv:1004.1776 [hep-th]].

\bibitem{Ade:2015ava} 
  P.~A.~R.~Ade {\it et al.} [Planck Collaboration],
  arXiv:1502.01592 [astro-ph.CO].

\bibitem{Page:1993wv} 
  D.~N.~Page,
  Phys.\ Rev.\ Lett.\  {\bf 71}, 3743 (1993)
  doi:10.1103/PhysRevLett.71.3743
  [hep-th/9306083].

\bibitem{Page:1993df} 
  D.~N.~Page,
  Phys.\ Rev.\ Lett.\  {\bf 71}, 1291 (1993)
  doi:10.1103/PhysRevLett.71.1291
  [gr-qc/9305007].


\end{thebibliography}
\end{document}